\documentclass[a4paper, 11pt]{article}
\usepackage{amsmath}
\usepackage{graphicx}
\usepackage{latexsym}

\def\b{\begin{eqnarray}}
\def\e{\end{eqnarray}}
\def\n{\noindent}

\begin{document}

\begin{center}

{\LARGE\textbf{Water waves and integrability \\}} \vspace {10mm}
\vspace{1mm} \noindent

{\large \bf Rossen I. Ivanov$^\ast$}\footnote{On leave from the
Institute for Nuclear Research and Nuclear Energy, Bulgarian
Academy of Sciences, Sofia, Bulgaria.} \vskip1cm \n \hskip-.3cm
\begin{tabular}{c}
\hskip-1cm $\phantom{R^R}${\it School of Mathematics, Trinity
College,}
\\ {\it Dublin 2, Ireland} \\
\\ {\it $^\ast$e-mail: ivanovr@maths.tcd.ie} \\
\\
\hskip-.8cm
\end{tabular}
\vskip1cm
\end{center}


\begin{abstract}

\n The Euler's equations describe the motion of inviscid fluid. In
the case of shallow water, when a perturbative asymtotic expansion
of the Euler's equations is taken (to a certain order of smallness
of the scale parameters), relations to certain integrable
equations emerge. Some recent results concerning the use of
integrable equation in modeling the motion of shallow water waves
are reviewed in this contribution.


\end{abstract}

\section{Governing equations for the inviscid fluid motion}
The motion of inviscid fluid with a constant density $\rho$ is
described by the Euler's equations: \b\label{eq1a} \frac{\partial
{\bf v}}{\partial
t}+({\bf v}\cdot\nabla){\bf v}&=&-\frac{1}{\rho}\nabla P+ {\bf g},\\
\nabla \cdot {\bf v}&=&0, \label{eq1b}\e

\n where ${\bf v}(x,y,z,t)$ is the velocity of the fluid at the
point $(x,y,z)$ at the time $t$, $P$ is the pressure in the fluid,
${\bf g}=(0,0,-g)$ is the constant Earth's gravity acceleration.

Consider now a motion of a shallow water over a flat bottom, which
is located at $z=0$. We assume that the motion is in the
$x$-direction, and that the physical variables do not depend on $y$.
Let $h$ be the mean level of the water and let $\eta (x,t)$
describes the shape of the water surface, i.e. the deviation from
the average level. The pressure is \b\label{eq2} P=P_A+\rho g
(h-z)+p(x,z,t), \e where $P_A$ is the constant atmospheric pressure,
and $p$ is a pressure variable, measuring the deviation from the
hydrostatic pressure distribution.  On the surface $z=h+\eta$,
$P=P_A$ and therefore $p=\eta \rho g$. Taking ${\bf v}\equiv(u,0,w)$
we can write the kinematic condition on the surface as [e.g. see
(Johnson 1997)] \b \label{eq3} w=\frac{\partial \eta}{\partial t}+ u
\frac{\partial \eta}{\partial x} \qquad \text{on}\qquad z=h+\eta. \e
Finally, there is no horizontal velocity at the bottom, thus

\b \label{eq4} w=0 \qquad \text{on}\qquad z=0. \e The equations
(\ref{eq1a}) -- (\ref{eq4}) give the system

\b &\phantom{=}& u_t+uu_x+wu_z=-\frac{1}{\rho}p_x,\nonumber \\ &\phantom{=}& w_t+uw_x+ww_z=-\frac{1}{\rho}p_z,\nonumber\\
&\phantom{=}& u_x+w_z=0, \nonumber
\\ &\phantom{=}&w= \eta_t+ u \eta_x,\quad  p=\eta \rho g, \qquad \text{on}\qquad z=h+\eta,\nonumber \\
&\phantom{=}& w=0 \qquad \text{on}\qquad z=0. \nonumber \\
\label{S1} \e

Let us introduce now dimensionless parameters $\varepsilon=a/h$ and
$\delta=h/\lambda$, where $a$ is the typical amplitude of the wave
and $\lambda$ is the typical wavelength of the wave. Now we can
introduce dimensionless quantities, according to the magnitude of
the physical quantities, see (Johnson 1997, 2002) for details: \b
x\rightarrow\lambda x, \qquad z\rightarrow z h,\qquad t\rightarrow
\frac{\lambda}{\sqrt{g h}}t,\qquad \eta\rightarrow a\eta, \qquad
\nonumber \\ u\rightarrow \varepsilon \sqrt{g h} u, \qquad
w\rightarrow \varepsilon \delta \sqrt{g h} w, \qquad p\rightarrow
\varepsilon \rho g h. \nonumber \e

This scaling is due to the observation that both $w$ and $p$ are
proportional to $\varepsilon$ i.e. the wave amplitude, since at
undisturbed surface ($\varepsilon=0$) both $w=0$ and $p=0$. The
system (\ref{S1}) in the new, dimensionless variables is

\b &\phantom{=}& u_t+\varepsilon (uu_x+wu_z)=-p_x,\nonumber \\ &\phantom{=}& \delta^2(w_t+\varepsilon (uw_x+ww_z))=-p_z,\nonumber\\
&\phantom{=}& u_x+w_z=0, \nonumber
\\ &\phantom{=}&w= \eta_t+ \varepsilon u \eta_x,\quad  p=\eta, \qquad \text{on}\qquad z=1+\varepsilon \eta,\nonumber \\
&\phantom{=}& w=0 \qquad \text{on}\qquad z=0. \nonumber \\
\label{S2} \e

For the right-running waves one can introduce the so-called {\it far
field} quantities, see (Johnson 1997, 2002, 2003$a$)

\b \label{eq5} \zeta=\sqrt{\varepsilon}(x-t), \qquad
\tau=\varepsilon^{3/2}t, \qquad w=\sqrt{\varepsilon}W,\e

\n and the system (\ref{S2}) acquires the form
\b &\phantom{=}& \varepsilon u_{\tau}-u_{\zeta}+\varepsilon (uu_{\zeta}+Wu_z)=-p_{\zeta},\label{eq6} \\
&\phantom{=}& \varepsilon \delta^2(\varepsilon W_{\tau}-W_{\zeta}+\varepsilon (uW_{\zeta}+WW_z))=-p_z,\label{eq7}\\
&\phantom{=}& u_{\zeta}+W_z=0, \label{eq8}
\\ &\phantom{=}&W= \varepsilon\eta_{\tau}-\eta_{\zeta}+ \varepsilon u \eta_{\zeta},\quad  p=\eta, \qquad \text{on}\qquad z=1+\varepsilon \eta,\label{eq9} \\
&\phantom{=}& W=0 \qquad \text{on}\qquad z=0. \label{eq10} \e

\section{Asymptotic expansion of the variables}

Following the idea of Johnson (2002), we can express the variables
$u$, $W$, $p$ as double-asymptotic expansion (in $\varepsilon$ and
$\delta$) with terms, depending only on $\eta(x,t)$ and explicitly
on $z$. As a result, a single nonlinear equation for $\eta$ will be
obtained, and thus all variables will be expressed through the
solution of this equation.

From (\ref{eq7}) it is evident that $p_z=O(\varepsilon \delta^2)$,
and thus in the leading order $p$ does not depend on $z$, i.e.
\b\label{eq11} p=\eta. \e Substitution of (\ref{eq11}) into
(\ref{eq6}) and (\ref{eq8}) gives for the leading orders
\b\label{eq12} u=\eta, \qquad W=-z\eta_{\zeta}. \e

Consider the next terms (first corrections) in the expansion of $u$
and $W$, denoted by $E(u)$, $E(W)$, which possibly contain terms of
orders $\varepsilon$ and $\delta^2$. Writing $u=\eta+E(u)$,
$W=-z\eta_{\zeta}+E(W)$, from (\ref{eq6}) it follows

\b\label{eq13} E_{\zeta}(u)=\varepsilon(\eta_{\tau}
+\eta\eta_{\zeta}),\e

\n and from (\ref{eq8}), (\ref{eq10}) and  (\ref{eq13}),

\b E(W)&=&-\int E_{\zeta}(u) {\text d} z =-\varepsilon z
(\eta_{\tau} +\eta \eta_{\zeta}),\nonumber \\\label{eq14}
W&=&-z(\eta_{\zeta}+\varepsilon \eta_{\tau} +\varepsilon \eta
\eta_{\zeta}).\e

Now the substitution of (\ref{eq14}) into (\ref{eq9}) gives the
leading order equation for $\eta$:

\b\label{eq15} \eta_{\tau}
=-\frac{3}{2}\eta\eta_{\zeta}+O(\varepsilon, \delta^2).\e

\n From (\ref{eq15}) and (\ref{eq13}) we obtain $E(u)=-\varepsilon
\eta^2/4$, i.e. no $\delta^2$ term is present and finally, using
(\ref{eq14}) and (\ref{eq15}),

\b\label{eq16} u=\eta-\frac{\varepsilon}{4}\eta^2, \qquad
W=-z\Big(\eta_{\zeta}-\frac{\varepsilon}{2}\eta\eta_{\zeta}\Big). \e

Using (\ref{eq12}) in (\ref{eq7}) we have $p_z=-\varepsilon \delta^2
z \eta_{\zeta \zeta}$. This can be integrated due to (\ref{eq9}) and
thus we obtain the next order approximation for $p$: \b\label{eq17}
p=\eta-\varepsilon \delta^2 \frac{1-z^2}{2} \eta_{\zeta \zeta}. \e

 We accomplished the first step, i.e. starting from the
leading order (\ref{eq11}), (\ref{eq12}), we obtained the first
corrections (\ref{eq16}), (\ref{eq17}) and an equation for $\eta$,
(\ref{eq15}). The next step can be performed in a similar fashion
and it gives

\b\label{eq18}
u&=&\eta-\frac{\varepsilon}{4}\eta^2+\frac{\varepsilon
^2}{8}\eta^3+\varepsilon
\delta^2\Big(\frac{1}{3}-\frac{z^2}{2}\Big)\eta_{\zeta \zeta},
\\\label{eq19}
W&=&-z\Big(\eta_{\zeta}-\frac{\varepsilon}{2}\eta\eta_{\zeta}+\frac{3\varepsilon^2}{8}\eta^2\eta_{\zeta}\Big)+\varepsilon
\delta^2 \Big(-\frac{z}{3}+\frac{z^3}{6}\Big)\eta_{\zeta \zeta
\zeta},
\\ \label{eq20}p&=&\eta-\varepsilon \delta^2 \frac{1-z^2}{2} \eta_{\zeta
\zeta}+\varepsilon^2 \delta^2 (\eta \eta_{\zeta \zeta}+(1-z^2)\eta^2
_{\zeta}), \e
 \n where $\eta(x,t)$ satisfies the equation
\b\label{eq21} \eta_{\tau}
=-\frac{3}{2}\eta\eta_{\zeta}+\frac{3}{8}\varepsilon\eta^2
\eta_{\zeta}-\frac{1}{6} \delta^2 \eta_{\zeta \zeta
\zeta}+O(\varepsilon^2, \delta^4, \varepsilon\delta^2).\e

\n We observe, that at the end of each step the equation for
$\eta(x,t)$ contains terms of smaller order than those, which appear
in the expressions for $u$ and $W$. Since we need an equation,
containing terms of order $O(\varepsilon^2, \delta^4,
\varepsilon\delta^2)$, we need to perform several intermediate
sub-steps, (like (\ref{eq13}), (\ref{eq14})) of the next step, which
leads to the desired equation

\b \eta_{\tau}
=-\frac{3}{2}\eta\eta_{\zeta}+\frac{3}{8}\varepsilon\eta^2
\eta_{\zeta}-\frac{3}{16}\varepsilon^2\eta^3
\eta_{\zeta}-\frac{1}{6} \delta^2 \eta_{\zeta \zeta \zeta} \phantom{******}\nonumber \\
-\frac{1}{24}\varepsilon\delta^2(23\eta_{\zeta}\eta_{\zeta\zeta}+10\eta
\eta_{\zeta\zeta \zeta})+O(\varepsilon^3, \delta^6,
\varepsilon^2\delta^2,\varepsilon\delta^4).\label{eq22}\e

\n Now, we can invert (\ref{eq18}) by specifying $u$ at a specific
depth, $z_0$

\n ($0\leq z_0 \leq 1$): defining $\hat{u}=u(\zeta,\tau,z_0)$, we
obtain

\b  \eta= \hat{u}+\frac{\varepsilon}{4}\hat{u}^2-\varepsilon
\delta^2\lambda \hat{u}_{\zeta
\zeta}+O(\varepsilon^3,\delta^6,\varepsilon \delta^4, \varepsilon^2
\delta^2),\label{eq23}\e where \b
\lambda\equiv\frac{1}{3}-\frac{z_0^2}{2},\qquad -\frac{1}{6}\leq
\lambda \leq\frac{1}{3}. \label{eq24}\e

Note that in (\ref{eq23}) there is no term of order $\varepsilon^2$.
The substitution of (\ref{eq23}) in (\ref{eq22}) yields: \b
\hat{u}_{\tau} =-\frac{3}{2}\hat{u}\hat{u}_{\zeta}-\frac{1}{6}
\delta^2 \hat{u}_{\zeta \zeta \zeta}
-\frac{1}{2}\varepsilon\delta^2\Big[\Big(\frac{29}{12}+6\lambda\Big)\hat{u}_{\zeta}\hat{u}_{\zeta\zeta}
+\frac{5}{6}\hat{u}\hat{u}_{\zeta\zeta
\zeta}\Big]\phantom{***}\nonumber
\\ +O(\varepsilon^3, \delta^6,
\varepsilon^2\delta^2,\varepsilon\delta^4).\label{eq25}\e

\n Next, we go back to the original variables, introducing
$T\equiv\sqrt{\varepsilon}t$, $X\equiv\sqrt{\varepsilon}x$, see
(\ref{eq5}), keeping only the scaling with $\varepsilon$ : \b
T=\frac{1}{\varepsilon}\tau, \qquad
X=\frac{1}{\varepsilon}\tau+\zeta, \label{eq26}\e

\n or $\partial_ \zeta=\partial_X$, $ \varepsilon \partial_
\tau=\partial_T+\partial_X.$
 \n Thus, (\ref{eq25}) yields

\b \hat{u}_T =-\hat{u}_X-\frac{3}{2}\varepsilon
\hat{u}\hat{u}_X-\frac{1}{6} \varepsilon \delta^2 \hat{u}_{XXX}
-\frac{1}{2}\varepsilon^2\delta^2\Big[\Big(\frac{29}{12}\!+\!6\lambda\Big)\hat{u}_{X}\hat{u}_{XX}
+\frac{5}{6}\hat{u}\hat{u}_{XXX}\Big]\nonumber
\\ +O(\varepsilon^4, \varepsilon \delta^6,
\varepsilon^3\delta^2,\varepsilon^2\delta^4).\phantom{*}\label{eq28}\e

\n Further, we add formally $(\varepsilon \delta^2 \mu
\hat{u}_{XXT}-\varepsilon \delta^2 \mu \hat{u}_{XXT})/2$ to the
left-hand side of (\ref{eq28}), where $\mu$ is an arbitrary real
parameter. In the first term we substitute $\hat{u}_{T}
=-\hat{u}_X-\frac{3}{2}\varepsilon \hat{u}\hat{u}_X$, according to
(\ref{eq28}):

 \b \Big(\hat{u}-\frac{1}{2}\varepsilon \delta^2 \mu \hat{u}_{XX}\Big)_T =-\hat{u}_X-\frac{3}{2}\varepsilon
\hat{u}\hat{u}_X+ \varepsilon \delta^2
(\frac{1}{2}\mu-\frac{1}{6})\hat{u}_{XXX}\phantom{********}\nonumber \\
-\frac{1}{2}\varepsilon^2\delta^2\Big[\Big(\frac{29}{12}+6\lambda-\frac{9}{2}\mu\Big)\hat{u}_{X}\hat{u}_{XX}
+\Big(\frac{5}{6}-\frac{3}{2}\mu\Big)\hat{u}\hat{u}_{XXX}\Big]\nonumber
\\ +O(\varepsilon^4, \varepsilon \delta^6,
\varepsilon^3\delta^2,\varepsilon^2\delta^4).\label{eq29}\e

\n We observe that (\ref{eq28}), (\ref{eq29}) do not contain terms
of orders $\varepsilon$ and $\varepsilon^2$. Thus, the set-up from
(Johnson 2002) naturally leads to the conclusion, that equations,
containing nonlinearities as those, appearing in the equations of
Camassa $\&$ Holm (1993), Fokas $\&$ Fuchssteiner (1981) [called
also CH from now on] and Degasperis $\&$ Procesi (1999),  Degasperis
{\it et al.} (2002) [DP for short], are generalizations of the
Korteweg-de Vries equation, containing the next order term
($\varepsilon^2\delta^2$) in the expansion with respect to the small
parameters $\varepsilon$, $\delta$.

\section{Integrable nonlinear equations}

In this section we start from a known integrable equation and we try
to write it in a form, in which it matches (\ref{eq29}) or
(\ref{eq28}). For another approach for matching between water waves
equations and integrable equations see (Dullin {\it et al.} 2003,
2004). The CH and DP equations can be written as


\b (U-U_{XX})_T=\omega U_X-(b+1)UU_X+bU_XU_{XX}+UU_{XXX},
\label{eq30}\e

\n where $U=U(X,T)$, $\omega$ is an arbitrary constant, $b=2$ for CH
and $b=3$ for DP. There is no other choice of the constant
coefficients in front of the nonlinear terms, leading to integrable
equations, see (Ivanov 2005$a$). Let us change the variables in
(\ref{eq30}) according to


\b X\rightarrow X-vT, \qquad T\rightarrow T, \qquad U\rightarrow
U+C,\label{eq31} \e

\n where $v$ and $C$ are arbitrary constants. Then (\ref{eq30})
acquires the form

\b(U-U_{XX})_T=[\omega-(b+1)C+v]U_X+(C-v)U_{XXX}\phantom{******}\nonumber
\\ -(b+1)UU_X+bU_XU_{XX}+UU_{XXX}. \label{eq32}\e

\n It is now clear that via the transforms (\ref{eq31}) one can
achieve arbitrary coefficients for the linear terms $U_X$ and
$U_{XXX}$. Let us now consider the following scaling of the
variables:


\b X\rightarrow \frac{1}{\alpha}X, \qquad T\rightarrow \beta T,
\qquad U\rightarrow \gamma U. \label{eq33} \e

Then (\ref{eq32}) can be written as \b(U-\alpha ^2 U_{XX})_T=\Gamma
_1 U_X+ \Gamma_2 U_{XXX}-\alpha\beta\gamma
(b+1)UU_X\phantom{*****}\nonumber
\\ +\alpha^{3}\beta\gamma(bU_XU_{XX}+UU_{XXX}), \label{eq34}\e

\n where $\Gamma_{1,2}$ are arbitrary constants. In order to match
(\ref{eq34}) to (\ref{eq29}) up to the given order, we need to make
the following identifications:

\b \alpha^2=\frac{1}{2}\varepsilon \delta^2\mu, \qquad
\alpha\beta\gamma (b+1)&=&\frac{3}{2}\varepsilon, \qquad
\alpha^{3}\beta\gamma=-\frac{1}{2}\Big(\frac{5}{6}-\frac{3}{2}\mu
\Big)\varepsilon^2\delta^2, \nonumber \\
\frac{29}{12}+6\lambda-\frac{9}{2}\mu&=&b\Big(\frac{5}{6}-\frac{3}{2}\mu
\Big)
 \label{eq35} \e

\n which are compatible iff

\b \mu=\frac{5(b+1)}{9b}, \qquad \lambda=\frac{30-9b}{72b}.
\label{eq36} \e

\n Thus, (\ref{eq36}) and (\ref{eq24}) show that (\ref{eq34})
describes water waves at depth

\b z_0=\sqrt{\frac{11b-10}{12b}}, \label{eq37} \e

\n i.e. CH ($b=2$) corresponds to $z_0=\frac{1}{\sqrt{2}}\approx
0.71$ and DP ($b=3$) corresponds to $z_0=\sqrt{\frac{23}{36}}\approx
0.80$. The scaling coefficients from (\ref{eq35}) are

\b \alpha=\sqrt{\frac{\mu}{2}}\varepsilon^{1/2}\delta, \qquad
\beta\gamma=\frac{3}{\sqrt{2\mu}(b+1)}\varepsilon^{1/2}\delta^{-1},
\label{eq38} \e

\n and, apparently only the product $\beta\gamma$ is determined,
i.e. there is additional freedom in the choice of $\beta$ and
$\gamma$, one can take, for simplicity, just $\gamma=1$ and then,
finally,

\b \alpha=\sqrt{\frac{\mu}{2}}\varepsilon^{1/2}\delta, \qquad
\beta=\frac{3}{\sqrt{2\mu}(b+1)}\varepsilon^{1/2}\delta^{-1}, \qquad
\gamma=1. \label{eq39}\e

Another equation, which passes the integrability check developed in
(Mikhailov $\&$ Novikov 2002; Sanders $\&$ Jing Ping Wang 1998;
Olver $\&$ Jing Ping Wang 2000) and is presumably integrable
(although we do not have a proof of this fact -- the test provides
only a necessary condition for integrability) is

\b(U-U_{XX}+U_{XXXX})_T=\Gamma_1 U_X-\Gamma_2U_{XXX}+\Gamma_2
U_{XXXXX}\phantom{******}\nonumber
\\ -UU_X+U_XU_{XX}-U_XU_{XXXX}, \label{eq40}\e

\n where $\Gamma_{1,2}$ are arbitrary constants. This equation
contains nonlinearities, similar to those, appearing in the
nonintegrable equations studied by Holm $\&$ Hone (2003).

The scaling (\ref{eq33}) gives \b(U-\alpha ^2
U_{XX}+\alpha^4U_{4X})_T&=&\alpha\beta\Gamma _1 U_X-\alpha^3\beta
\Gamma_2 U_{XXX}+\alpha^5\beta \Gamma_2 U_{5X}\nonumber
\\ \!\!\!&-&\!\!\!\alpha\beta\gamma UU_X+\alpha^{3}\beta\gamma U_XU_{XX}-\alpha^{5}\beta\gamma U_XU_{4X},\nonumber\\ \label{eq41}\e

\n The matching between (\ref{eq41}) and (\ref{eq29}) leads to the
following identifications:

\b \alpha^2&=&\frac{1}{2}\varepsilon \delta^2\mu, \qquad
\alpha\beta\gamma =\frac{3}{2}\varepsilon, \qquad
\alpha^{3}\beta\gamma=-\frac{1}{2}\Big(6\lambda-\frac{9}{2}\mu+\frac{29}{12}\mu
\Big)\varepsilon^2\delta^2, \nonumber \\
\mu&=&\frac{5}{9}. \label{eq42} \e

\n In a similar way, from (\ref{eq41}) we find (assuming again
$\gamma=1$)

\b \lambda=-\frac{1}{8}, \qquad z_0=
\sqrt{\frac{11}{12}}\approx0.96; \label{eq43} \e

\b \alpha=\sqrt{\frac{\mu}{2}}\varepsilon^{1/2}\delta, \qquad
\beta=\frac{3}{\sqrt{2\mu}}\varepsilon^{1/2}\delta^{-1}, \qquad
(\gamma=1). \label{eq44}\e

\n The terms with fourth and fifth derivative in (\ref{eq41}) are of
orders

\b \alpha^4=\frac{\mu^2}{4}\varepsilon^2\delta^4, \qquad
\alpha^5\beta=\frac{3\mu^2}{8}\varepsilon^{3}\delta^{4}, \qquad
\alpha^5\beta\gamma=\frac{3\mu^2}{8}\varepsilon^{3}\delta^{4},
\label{eq45}\e

\n and, therefore are small, in comparison to the other terms.

Another set of integrable equations is of the type

\b U_T+U_{XXXXX}+2(6b+1)U_X
U_{XX}+4(b+1)UU_{XXX}+20bU^2U_X=0,\nonumber\\ \label{eq46}\e

\n where one can recover the Caudrey-Dodd-Gibbon equation, (Caudrey
{\it et al.} 1976) for $b=1/4$, the Sawada-Kotera equation, (Sawada
$\&$ Kotera 1974) for $b=3/2$ and the Kaup-Kuperschmidt equation,
(Kaup 1980) for $b=4$.

It is a natural question to ask, if (\ref{eq46}) can match
(\ref{eq28}). Applying the transformation (\ref{eq31}) to
(\ref{eq46}), we can write it in the form

\b U_T=\Gamma U_X-4(b+1)CU_{XXX}-U_{5X}-40bCUU_X \phantom{*********}\nonumber\\
-2(6b+1)U_X U_{XX}-4(b+1)UU_{XXX}-20bU^2U_X, \label{eq47}\e

\n where $\Gamma=v+20bC^2$ can apparently be arranged to be an
arbitrary constant with the help of the free parameter $v$. The
scaling (\ref{eq33}), applied to (\ref{eq47}) gives:

\b U_T=\alpha\beta\Gamma U_X-4(b+1)C\alpha^3\beta U_{XXX}-\alpha^5 U_{5X}-40bC\alpha\beta\gamma UU_X \phantom{***}\nonumber\\
-\alpha^3\beta\gamma[2(6b+1)U_X
U_{XX}+4(b+1)UU_{XXX}]-20b\alpha\beta\gamma^2U^2U_X. \label{eq47a}\e
Apparently we need to make the following identifications:

\b 4(b+1)C\alpha^3\beta=\frac{1}{6}\varepsilon \delta^2, \quad 40bC
\alpha\beta\gamma =\frac{3}{2}\varepsilon, \quad
4(b+1)\alpha^{3}\beta\gamma=\frac{5}{12}\varepsilon^2 \delta^2,
 \label{eq48} \e

 \n giving

 \b \alpha\beta=\frac{3}{200 bC^2}, \quad \gamma
 =\frac{5C}{2}\varepsilon.
 \label{eq49} \e

The order of the term $U^2U_X$ is
$20b\alpha\beta\gamma^2=\frac{3}{16}\varepsilon^2$, it is not small
in comparison to the other terms, and therefore cannot be neglected.
Thus, there is no direct match between (\ref{eq46}) and
(\ref{eq28}), however, there is more complicated transformation,
given in (Fokas $\&$ Liu 1996) [based on the Kodama transform,
(Kodama 1985)] providing the link between the water-wave equations
and the integrable systems (\ref{eq46}).

\section{Water waves moving over a shear flow}

So far we have only considered waves in the absence of shear. Now
let us notice that there is an exact solution of the governing
equations (\ref{S1}) of the form $u=\tilde{U}(z)$, $0\leq z\leq h$,
$w\equiv 0$, $p\equiv 0$, $\eta \equiv 0$. This solution is nothing,
but an arbitrary underlying 'shear' flow. Waves of small amplitude
(of order $\varepsilon$) propagating over this underlying flow are
studied by many authors and here we will partially follow Johnson
(2003$a$) and Burns (1953). The scaling for such solution is clearly
\b u\rightarrow \sqrt{g h}\Big(\tilde{U}(z)+\varepsilon
u\Big),\nonumber \e and the scaling for the other variables is as
before. Thus, from (\ref{S1}) instead of (\ref{S2}) in this case we
have

\b &\phantom{=}& u_t+\tilde{U}u_x+w\tilde{U}'+\varepsilon (uu_x+wu_z)=-p_x,\nonumber \\
&\phantom{=}& \delta^2(w_t+\tilde{U}w_x+\varepsilon (uw_x+ww_z))=-p_z,\nonumber\\
&\phantom{=}& u_x+w_z=0, \nonumber
\\ &\phantom{=}&w= \eta_t+ (\tilde{U}+\varepsilon u )\eta_x,\quad  p=\eta, \qquad \text{on}\qquad z=1+\varepsilon \eta,\nonumber \\
&\phantom{=}& w=0 \qquad \text{on}\qquad z=0. \nonumber \\
\label{S22} \e The prime denotes derivative with respect to $z$.

In what follows we need the propagation speed $c$ of the waves in
the linear approximation, i.e. in the case when in (\ref{S22}) it
is taken $\varepsilon=\delta=0$. This velocity is now not
independent on $\tilde{U}$.  Since $p_z=0$, in the linear
approximation $p=\eta$.  Let us introduce a stream function
$\psi$, such that $u=\psi_z$ and $w=-\psi_x$. In the case of
linear waves we can assume that $\psi=\phi(z)e^{i k (x-ct)}$,
$\eta=\eta_0 e^{i k (x-ct)}$, where $k$ is a wave number and
$\eta_0$ is a constant. From (\ref{S22}) with
$\varepsilon=\delta=0$ we now easily find a relation between
$\tilde{U}(z)$ and $\phi(z)$:

\b\label{eq50} \phi'(\tilde{U}-c)-\tilde{U}'\phi+\eta_0=0, \qquad
\phi(1)=-(U(1)-c)\eta_0, \qquad \phi(0)=0.\e

\n The first equation in (\ref{eq50}) can be written as \b
\frac{{\text d}}{{\text d}
z}\frac{\phi}{\tilde{U}-c}=-\frac{\eta_0}{(\tilde{U}-c)^2},\nonumber\e
and can be integrated directly. Imposing the boundary conditions
from (\ref{eq50}) we finally obtain the following relation for the
speed of propagation $c$ (the so-called Burns condition):

\b\label{eq51} \int_0^1 \frac{{\text d}z}{[\tilde{U}(z)-c]^2}=1.\e

For a nondecreasing function $\tilde{U}(z)$, such that
$\tilde{U}(0)\leq \tilde{U}(z)\leq \tilde{U}(1)$ there are always
two solutions: $c>\tilde{U}(1)$ and $c<\tilde{U}(0)$. In the absence
of flow, $\tilde{U}\equiv 0$ these two solutions are simply $c=\pm
1$. The presentation in the previous sections corresponds to the
choice $c=1$.

Again, we introduce the far field variables, cf. (\ref{eq5})

\b \label{eq52} \zeta=\sqrt{\varepsilon}(x-ct), \qquad
\tau=\varepsilon^{3/2}t, \qquad w=\sqrt{\varepsilon}W,\e

\n and the system (\ref{S22}) acquires the form
\b &\phantom{=}& \varepsilon u_{\tau}+(\tilde{U}-c)u_{\zeta}+W\tilde{U}'+\varepsilon (uu_{\zeta}+Wu_z)=-p_{\zeta},\nonumber \\
&\phantom{=}& \varepsilon \delta^2(\varepsilon W_{\tau}+(\tilde{U}-c)W_{\zeta}+\varepsilon (uW_{\zeta}+WW_z))=-p_z,\nonumber\\
&\phantom{=}& u_{\zeta}+W_z=0, \nonumber
\\ &\phantom{=}&W= \varepsilon\eta_{\tau}+(\tilde{U}-c)\eta_{\zeta}+ \varepsilon u \eta_{\zeta},\quad  p=\eta, \qquad \text{on}\qquad z=1+\varepsilon \eta,\nonumber \\
&\phantom{=}& W=0 \qquad \text{on}\qquad z=0. \nonumber
\\ \label{S32} \e

Now let us concentrate to the simplest nontrivial case: a linear
shear, $\tilde{U}(z)=Az$, where $A$ is a constant. We choose $A>0$,
so that the underlying flow is propagating in the positive direction
of the $x$-coordinate. The condition (\ref{eq51}) gives the
following expression for $c$: \b\label{eq53}
c=\frac{1}{2}\Big(A\pm\sqrt{4+A^2}\Big).\e If there is no shear
($A=0$), then $c=\pm 1$.

Of course, a parabolic distribution
$\tilde{U}(z)=\tilde{U}(1)(2z-z^2)$ would be more realistic, but
then the solution of (\ref{eq51}) is not so simple, cf. (Burns
1953).

The solution of the system (\ref{S32}) can be obtained as a series
in $\varepsilon$ and $\delta$ following the method, explained in the
previous section. Here we present the final result, obtained in
Johnson (2003$a$). The equation for $\eta$ is \b \eta_{\tau}
=-\frac{c^4+c^2+1}{c(c^2+1)}\eta\eta_{\zeta}+\frac{c(c^4+4c^2+1)}{2(c^2+1)^3}\varepsilon\eta^2
\eta_{\zeta}-\frac{1}{3c(c^2+1)} \delta^2 \eta_{\zeta \zeta \zeta} \phantom{******}\nonumber \\
-\frac{1}{3c(c^2+1)^3}\varepsilon\delta^2[(2c^6+4c^4+11c^2+6)\eta_{\zeta}\eta_{\zeta\zeta}+(c^4+6c^2+3)\eta
\eta_{\zeta\zeta \zeta}]\nonumber \\
+O(\varepsilon^2, \delta^4).\label{eq54}\e

In the no-shear case  and right-going waves ($c=1$) one can recover
(\ref{eq22}); $c=-1$ corresponds to left-going waves. The horizontal
velocity to this order is

\b\label{eq55} u=\frac{1}{c}\eta-\varepsilon\frac{
c}{2(c^2+1)}\eta^2+\varepsilon
\delta^2\Big[\frac{c^2+3}{6c(c^2+1)}-\frac{z^2}{2c}\Big]\eta_{\zeta
\zeta}. \e

\n Again, in order to invert (\ref{eq54}) we have to specify $u$ at
a specific depth, $z_0$ ($0\leq z_0 \leq 1$):
$\hat{u}=u(\zeta,\tau,z_0)$. The result is

\b  \eta=
c\hat{u}+\varepsilon\frac{c^4}{2(c^2+1)}\hat{u}^2-\varepsilon
\delta^2c^2\Lambda \hat{u}_{\zeta \zeta},\quad \text{where}\quad
\Lambda\equiv\frac{3+c^2}{6c(1+c^2)}-\frac{z_0^2}{2c}.\label{eq56}\e

It is convenient to introduce a new dependent variable

\b  V=
\hat{u}+\varepsilon\sigma\hat{u}^2+O(\varepsilon^2,\delta^4),\qquad
\sigma\equiv \frac{c^5(c^4+c^2-2)}{2(c^4+c^2+1)(c^2+1)^2}
\label{eq58}\e (which is NOT the Kodama transform) for which the
equation is

\b  V_{\tau} =-\frac{c^4+c^2+1}{(c^2+1)}VV_{\zeta}-\delta^2\frac{1}{3c(c^2+1)} V_{\zeta \zeta \zeta} \phantom{****************}\nonumber \\
-\varepsilon\delta^2\Big\{\Big[2c\Lambda
\frac{c^4+c^2+1}{c^2+1}+\frac{2c^6+7c^4+14c^2+6}{3(c^2+1)^3}-\frac{2\sigma}{c(c^2+1)}\Big]V_{\zeta}V_{\zeta\zeta}\nonumber \\
+\frac{c^4+6c^2+3}{3(c^2+1)^3}VV_{\zeta\zeta
\zeta}\Big\}+O(\varepsilon^2, \delta^4).\label{eq59}\e

Next, we return to the original variables in (\ref{eq59}), up to a
scaling: $T\equiv\sqrt{\varepsilon}t$, $X\equiv\sqrt{\varepsilon}x$,
see (\ref{eq52}):  $T=\varepsilon ^{-1}\tau$, $X=\zeta
+\varepsilon^{-1}c \tau$,  then  $ \partial _{\zeta}=\partial_X$;
$\varepsilon \partial_{ \tau}=\partial_ T+c\partial_X$.

In conjunction, we add  $(\varepsilon \delta^2 \mu
V_{XXT}-\varepsilon \delta^2 \mu V_{XXT})/(1+c^2)$ where $\mu$ is an
arbitrary real parameter. In the first term we substitute the
leading order of \b V_{T} \sim -V_X-\varepsilon
\frac{c^4+c^2+1}{c^2+1}VV_X \nonumber \e and obtain

\b  V_{T} =-V_X-\varepsilon\frac{c^4+c^2+1}{c^2+1}VV_X-\varepsilon\delta^2\frac{1-3\mu c}{3c(c^2+1)} V_{XXX}
+\varepsilon\delta^2\frac{\mu }{c^2+1} V_{XXT}\phantom{*****}\nonumber \\
-\varepsilon^2\delta^2\Big[
\frac{[2c\Lambda(c^2\!+\!1)-3\mu](c^4\!+\!c^2\!+\!1)}{(c^2+1)^2}+\frac{2c^6\!+\!7c^4\!+\!14c^2\!+\!6}{3(c^2+1)^3}-\frac{2\sigma}{c(c^2\!+\!1)}\Big]V_XV_{XX}\nonumber \\
-\varepsilon^2\delta^2\Big[\frac{c^4+6c^2+3}{3(c^2+1)^3}-\mu\frac{c^4+c^2+1}{(c^2+1)^2}\Big]VV_{XXX}+O(\varepsilon^3,
\varepsilon\delta^4).\nonumber\e \b \label{eq62}\e

The comparison between (\ref{eq62}) and (\ref{eq34}) gives the
following possibilities for the parameters, cf.
(\ref{eq35})--(\ref{eq37}):

\b \mu&=&\frac{5(b+1)(c^4+6c^2+3)}{3b(c^2+1)(c^4+c^2+1)}, \nonumber\\
\Lambda&=&\frac{-2bc^{10}+(3-4b)c^8+(21-6b)c^6+(30-13b)c^4+(27-2b)c^2+9}{6bc(c^2+1)^2(c^4+c^2+1)^2}.
\nonumber \e

\n Therefore, according to the relation between $z_0$ and $\Lambda$
in (\ref{eq56}), for a given propagation speed $c$, (\ref{eq34})
describes water waves at depth

\b
z_0^2\!=\!\frac{bc^{12}+8bc^{10}+(18b\!-\!3)c^8+(26b\!-\!21)c^6+(31b\!-\!\!30)c^4+(12b\!-\!27)c^2+3b\!-\!9}{3b(c^2+1)^2(c^4+c^2+1)^2}.
\nonumber \e In the case of right-moving wave without underlying
flow ($c=1$) we recover (\ref{eq37}). For the CH equation ($b=2$)
this gives [cf. (Johnson 2003a)] \b
z_0=\frac{\sqrt{2c^{12}+16c^{10}+33c^8+31c^6+32c^4-3c^2-3}}{\sqrt{6}(c^2+1)(c^4+c^2+1)},
\label{eq63} \e \n and for the DP equation ($b=3$)\b
z_0=\frac{|c|\sqrt{c^{10}+8c^8+17c^6+19c^4+21c^2+3}}{\sqrt{3}(c^2+1)(c^4+c^2+1)}.
\label{eq64} \e

The comparison between (\ref{eq62}) and (\ref{eq41}) gives [cf.
(\ref{eq42})--(\ref{eq45})]

\b \mu&=&\frac{c^4+6c^2+3}{3(c^2+1)(c^4+c^2+1)}, \qquad 
\Lambda=-\frac{c^2(c^2+2)(2c^6+6c^2+1)}{6c(c^2+1)^2(c^4+c^2+1)^2},
\nonumber \\
z_0\!&=&\!\frac{\sqrt{c^{12}+8c^{10}+18c^8+26c^6+31c^4+12c^2+3}}{\sqrt{3}(c^2+1)(c^4+c^2+1)}.
\label{eq65} \e

\n As expected, $c=1$ in (\ref{eq65}) gives the result from
(\ref{eq43}), $z_0\approx 0.96$.

\begin{figure}
\centering
\includegraphics[height=7cm]{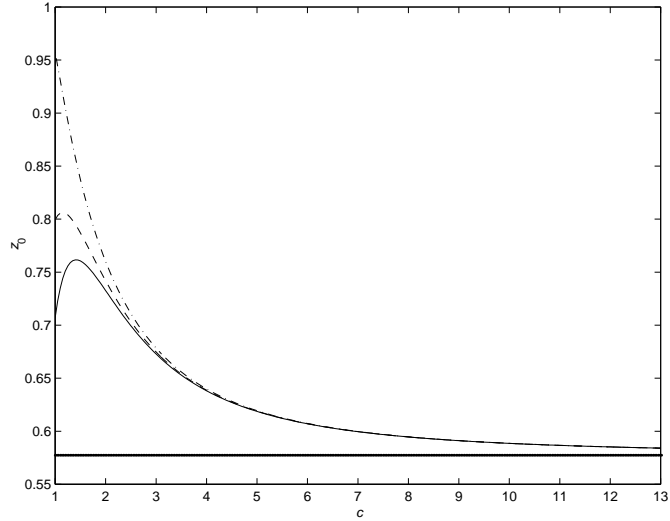}
%
%
\caption{ Downstream propagation: plot of the dependence of $z_0$ on
$c$. (\ref{eq63}) for CH -- solid line; (\ref{eq64}) for DP --
dashed line; (\ref{eq65}) for equation (\ref{eq41}) -- dash-dotted
line. All cases have a horizontal asymptote $z_0 \rightarrow
\frac{1}{\sqrt{3}}\approx 0.577$ as $c\rightarrow\infty$. The CH
graph has a maximum $z_0\approx 0.762$ at $c\approx 1.416$; The DP
graph has a maximum $z_0\approx 0.806$ at $c\approx 1.149$.}
\label{downstream}
\end{figure}

\begin{figure}
\centering
\includegraphics[height=7cm]{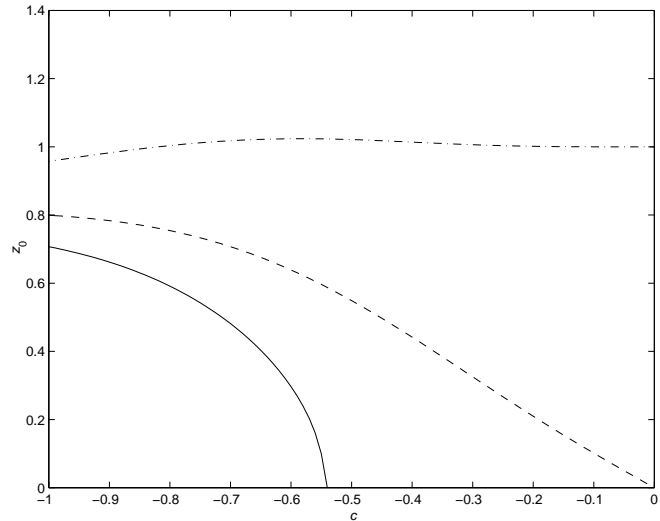}
%
%
\caption{ Upstream propagation: plot of the dependence of $z_0$ on
$c$. (\ref{eq63}) for CH -- solid line; (\ref{eq64}) for DP --
dashed line; (\ref{eq65}) for equation (\ref{eq41}) -- dash-dotted
line. The function (\ref{eq63}) is not real for $c_0\leq c\leq 0$,
$c_0\approx -0.544$; The graph of (\ref{eq65}) has a maximum
$z_0\approx 1.0237$ at $c\approx -0.600$.} \label{upstream}
\end{figure}

Without loss of generality we can assume that the underlying flow is
propagating in the positive direction of the $x$-coordinate, i.e.
$A>0$. Then (\ref{eq53}) leads to the following restriction to the
possible values of $c$: $\ c\geq1$ (for the waves, moving in the
direction of the flow, downstream) or $-1\leq c<0$ (for the waves,
propagating upstream). The plot of the dependence of $z_0$ on $c$ ,
according to the equations (\ref{eq63}), (\ref{eq64}) and
(\ref{eq65}) is given on Fig.\ref{downstream} and
Fig.\ref{upstream}. From Fig.\ref{upstream} we notice that there is
a region, where the function (\ref{eq63}) is not real: $c_0\leq
c\leq 0$, $c_0\approx -0.544$, and therefore the CH (in this
setting) is not a relevant model for these values of $c$. Although
the graph of (\ref{eq65}) has a maximum $z_0\approx 1.0237$ at
$c\approx -0.600$. which violates the condition $z_0\leq1$, we
notice that in the case of upstream propagation (Fig.\ref{upstream})
for all possible values of $c$ one can assume $z_0\approx 1$, i.e.
equation (\ref{eq40}) models very well the wave propagation on the
surface ($z_0=1$) for all possible velocities.

\section{Conclusions}
In conclusion, CH, DP and (\ref{eq40}) describe in a direct way
(without the use of the Kodama transform) the velocity (and,
consequently all related variables) of shallow water waves at depths
$0.71h$, $0.80h$ and $0.96h$ correspondingly (in the absence of a
shear flow), where $h$ is the depth of the undisturbed water. From a
modeling point of view the advantage of the CH and DP equations over
KdV consists  also in the fact that they capture the wave-breaking
phenomenon, cf. (Constantin 2000), (Constantin $\&$ Escher 1998) and
(Zhou 2004). These equations can also be used as water wave models
in the presence of an arbitrary shear flow. It is always more
convenient to work with integrable equations, since their solutions
are explicitly known or can be, in principle, explicitly
constructed. For CH the solitons are stable patterns and thus
physically recognizable, e.g. see the papers by Constantin $\&$
Strauss (2000, 2002), Constantin $\&$ Molinet (2001). The
$N$-soliton solution for CH is explicitly obtained by the Inverse
Scattering Method in (Constantin {\it et al.} 2006) [see also the
earlier works on the CH spectral problem by Constantin (1998, 2001);
Constantin $\&$ McKean, (1999)]. In parametric form the $N$-soliton
solution for CH is obtained by: Johnson (2003$b$), Li $\&$ Zhang
(2004), Li (2005), Parker (2004, 2005a, 2005b), Matsuno (2005$b$).
Other types of explicitly known CH solutions are: multi-peakons
(Beals {\it et al.}, 2003), periodic solutions (Gesztesy $\&$ Holden
2003), traveling-waves (Parkes $\&$ Vakhnenko 2005), (Lenells
2005$a$). The construction of multi-soliton and multi-positon
solutions for the Associated Camassa-Holm equation using the
Darboux/B{\"a}cklund transform is presented in (Schiff 1998), (Hone
1999) and (Ivanov, 2005b). The $N$-soliton solution for the DP
equation was recently derived by Matsuno (2005$a$), the multi-peakon
solutions for DP are obtained by Lundmark $\&$ Szmigielski (2003,
2005); the traveling waves -- by Parkes $\&$ Vakhnenko (2004) and
Lenells (2005$b$). There are similarities between the CH and DP
equations, in a sense that they both are integrable and have a
hydrodynamic derivation. However, it is interesting to notice that
only CH has a geometric interpretation as a geodesic flow, cf.
(Constantin $\&$ Kolev 2003), (Kolev 2004).

\section*{Acknowledgements} The author acknowledges funding from the
Irish Research Council for Science, Engineering and Technology.
This paper was written while the author participated in the
program "Wave Motion" at the Mittag-Leffler Institute, Stockholm,
in the Fall of 2005.

\label{lastpage}
\end{document}